\documentclass[nofootinbib]{revtex4}
\usepackage{amssymb}
\begin{document}
\title{Some aspects of the phenomenology of canonical
    noncommutative spacetimes\footnote{\uppercase{T}his work
is based on Refs. [1] and [2], to which \uppercase{I} refer the
reader for all details.}}
\author{\textbf{Luisa Doplicher}}
\affiliation{Dipart. Fisica, Univ.~Roma ``La Sapienza'', P.le Moro
2,
\\00185 Roma, Italy}
\email{doplicher@roma1.infn.it}

\begin{abstract}

I describe some phenomenological contexts in which it is possible
to investigate effects induced by (string-motivated) canonical
noncommutative spacetime. Due to the peculiar structure of the
theory the usual criteria adopted for the choice of experimental
contexts in which to test a theory may not be applicable here;
care is required in taking into account the effects of IR/UV
mixing. This invites one to consider contexts involving particles
of relatively high energies, like high-energy cosmic rays and
certain high-energy gamma rays observed from distant
astrophysical sources.

\end{abstract}

\maketitle

Canonical noncommutative spacetime is characterized by coordinate noncommutativity
of the form:
\begin{equation}
[x^\mu,x^\nu] = i \theta^{\mu \nu} \quad .
\end{equation}
An increase of interest in this type of noncommutativity
has been recently motivated by the observation that
it may be relevant to the description of some aspects of
string theories formulated in
presence of a background $B$ field.

Since they reflect the properties of a background the
$\theta^{\mu \nu}$ parameters cannot be observer independent;
thus, as a normal background field would, they break
\cite{1st,2nd,mst,carlson} some spacetime symmetries of the
theory. This symmetry loss causes one-loop $E(p)$ dispersion
relations to be deformed. Dispersion relation deformations in (non
supersymmetric) $U(1)$ theory turn out to be as follows. For
photons the modification to the self-energy is proportional to
\cite{mst}
\begin{equation}
\frac{\left(p \theta \right)^\mu \left(p \theta
\right)^\nu}{\left(p \theta \right)^4} \quad,
\end{equation}
(where $\left(p \theta \right)^\mu$ is shorthand for $p_\nu
\theta^{\mu \nu}$) and thus the deformed dispersion relation is
\begin{equation}
p_0^2 =\vec{p}^2 + \frac{C}{\left(p \theta \right)^2}
\label{photon}\quad;
\end{equation}
where $C$ is a constant. This dispersion relation deformation is
polarization dependent, in the sense that only photons with
polarization vector parallel to the direction of the vector
$\left(p \theta \right)^\mu$ are affected by the modification.
For neutral scalars the modification to the self-energy is
proportional to $\left(p \theta \right)^{-2}$ and thus the
dispersion relation is modified in this way:
\begin{equation}
p_0^2 =m^2+ \vec{p}^2 + \frac{C^\prime}{\left(p \theta \right)^2}
\label{scalar} \quad ,
\end{equation}
with $C^\prime$ a different constant. Finally, for Majorana
fermions the dispersion relation receives only logarithmic
corrections, mass suppressed.

A fundamental point are the energy scales suitable for
phenomenology. Of course it is necessary to introduce an UV cutoff
$\Lambda$, since the theory is of course not viewed as fundamental
(for example it does not include gravity). But through the IR/UV
mixing \cite{mvrs} the lack of knowledge of the UV sector (above
$\Lambda$) also implies a lack of knowledge of the structure of
the theory at scales lower than\footnote{For simplicity I am
assuming that the matrix $\theta^{\mu \nu}$ can be described in
terms of a single characteristic scale $\theta$.}
 $\mu=\left( \Lambda \theta \right)^{-1}$, which should therefore
be treated as an IR cutoff in phenomenology.

In presence of canonical noncommutativity our lack of control at
energy scales higher than a certain cutoff scale $\Lambda$
implies, because of the IR/UV mixing, also a lack of control at
energy scales below $\mu$.

This means that testing the theory at very low energies (below
the energy scale $\mu$) does not yield reliable results. Of
course in setting up a phenomenology that focuses on the regime
$\mu=\left( \Lambda \theta \right)^{-1} < E < \Lambda $ a key
issue is the value of $\mu$. And for some plausible choices of
cutoff $\Lambda$ and noncommutativity parameters one finds
\cite{1st} that $\mu$ can be rather high, possibly as high as
$\mu \sim TeV$. This implies that to test the theory one must
compare it with data corresponding to relatively high energies.
In such cases our best chances are clearly found in astrophysics.

The deformation (\ref{photon}) of the photon dispersion relation
brings about a deformation of the propagation of light; the speed
of light acquires a dependence upon wavelength. A good opportunity
to test this effect is provided by Blazars \cite{grbgac,aus},
which indeed have been observed \cite{billetal} at and above the
TeV scale. For some choices of the cutoff $\Lambda$ the analysis
of recently-observed Blazars allows \cite{1st} to set limits on
$\theta$ in the neighborhood of $\theta \gtrsim 10^{-40} cm^2$
The surprising fact that limit takes the form of an upper bound
reflects some features of the IR/UV mixing. In fact, the
observations only allow to determine that one of two options is
realized: either the cutoff estimate is incorrect or a suitable
upper bound on $\theta$ must be inforced. The mentioned
polarization dependence of the effect implies that, for
unpolarized sources, only a fraction of the flux can be used for
the analysis.

In the context of observations of cosmic rays an opportunity
for canonical noncommutativity to enter the analysis originates
from (\ref{scalar}), which suggests a modification
of the dispersion relation for pions.
This would affect the threshold energy requirements
for photopion production,
\begin{equation}
p + \gamma = p + \pi \quad,
\end{equation}
which is known to be relevant for the analysis of the interactions
between high-energy cosmic rays and CMB (cosmic microwave
background) photons. The usual special relativistic analysis of
photopion production leads to the prediction that the flux of
cosmic rays with energies above a cutoff energy of $10^{19}$ eV
should be strongly suppressed. The value of this cutoff energy
can be significantly higher \cite{1st} if one takes into account
the modified dispersion relation (\ref{scalar}) for the pion,
while keeping unmodified the law of energy-momentum conservation
and the dispersion relations for the proton and the CMB photon.
Support for an unmodified law of energy-momentum conservation is
found \cite{2nd,douglas} in the analysis of field theories in
canonical noncommutative spacetime\footnote{Note however that,
while in canonical noncommutativity it appears appropriate to
assume modified dispersion relations and unmodified laws of
energy-momentum conservation, there are other proposals for
quantum spacetime in which both a modification of the dispersion
relation and a modification of energy-momentum conservation are
present \cite{dsr}.}, the assumption of unmodified proton
dispersion relation is justified by some general properties
\cite{gaume} of charged spin-$1/2$ high-energy particles in
canonical noncommutativity, and we assumed \cite{1st} an
unmodified dispersion relation for CMB photons because of their
very low typical energies (likely below the mentioned low-energy
cutoff of validity of our framework) and the strong experimental
basis for an unmodified dispersion relation for low-energy
photons. Forthcoming cosmic-ray observatories could be sensitive
to the effects of canonical noncommutativity if $\theta$ takes a
value which is roughly in the neighborhood \cite{1st} of
$10^{-38}cm^2$.


\begin{thebibliography}{99}

\bibitem{1st}
G.~Amelino-Camelia, L.~Doplicher, S.~Nam and Y.~S.~Seo,
Phys.\ Rev.\ D {\bf 67} (2003) 085008, hep-th/0109191.

\bibitem{2nd}
G.~Amelino-Camelia, M.~Arzano and L.~Doplicher,
hep-th/0205047.

\bibitem{mst}
A.~Matusis, L.~Susskind and N.~Toumbas,
JHEP {\bf 0012} (2000) 002, hep-th/0002075.

\bibitem{carlson} C.E.~Carlson, C.D.~Carone and R.F.~Lebed,
Phys. Lett. B549:337-343,2002, hep-ph/0209077.

\bibitem{mvrs}
S.~Minwalla, M.~Van Raamsdonk and N.~Seiberg,
JHEP {\bf 0002} (2000) 020, hep-th/9912072.

\bibitem{grbgac} G. Amelino-Camelia, J. Ellis, N.E. Mavromatos,
D.V. Nanopoulos and S. Sarkar, Nature {393} (1998) 763,
astro-ph/9712103; G.~Amelino-Camelia and T.~Piran, Phys.~Rev.~D64
(2001) 036005, astro-ph/0008107.

\bibitem{aus} R.J.~Protheroe and H.~Meyer,
Phys.~Lett.~B493 (2000) 1, astro-ph/0005349.

\bibitem{billetal} S.D.~Biller {\it et al},
Phys.~Rev.~Lett.~83 (1999) 2108.

\bibitem{douglas} N.R.~Douglas and N.A.~Nekrasov,
Rev.~Mod.~Phys.~73 (2001) 977.

\bibitem{dsr} G.~Amelino-Camelia,
Int.~J.~Mod.~Phys.~D11 (2002) 35, gr-qc/0012051;
Phys.~Lett.~B510 (2001) 255, hep-th/0012238.

\bibitem{gaume} L.~Alvarez-Gaume and J.L.F.~Barbon,
Int.J.Mod.Phys.A16:1123-1146,2001, hep-th/0006209.


\end{thebibliography}
\end{document}